\documentclass[12pt]{article}
\usepackage{epsfig}
\usepackage{amsmath,lscape,sidecap,epsfig}
\usepackage{booktabs}
\usepackage{graphicx}
\usepackage{rotating}
\usepackage{amssymb}
\usepackage{color}
\usepackage{ulem}
\usepackage{tikz}
\usepackage{color}

\def\beq{\begin{equation}}
\def\eeq{\end{equation}}
\def\beqa{\begin{eqnarray}}
\def\eeqa{\end{eqnarray}}
\def\ban{\begin{eqnarray*}}
\def\ean{\end{eqnarray*}}
\def\bi{\begin{itemize}}
\def\ei{\end{itemize}}

\usepackage{ulem}

\newcommand{\unit}[1]{\ensuremath{\, \mathrm{#1}}}

\newcommand{\pc}[1]{\ensuremath{\left(#1\right)}}

\def\beq{\begin{equation}}
\def\eeq#1{\label{#1}\end{equation}}
\def\eeqn{\end{equation}}

\def\beqa{\begin{eqnarray}}
\def\eeqa#1{\label{#1}\end{eqnarray}}
\def\eeqan{\end{eqnarray}}

\let\bar=\overbar

\def\Dslash{\not{\hbox{\kern-4pt $D$}}}
\def\dslash{\not{\hbox{\kern-2pt $\del$}}}

\def\msb{{\bar{\ssstyle M \kern -1pt S}}}

\usepackage{fancyhdr,graphicx}
\def\Title#1{\begin{center} {\Large {\bf #1} } \end{center}}

\begin{document}

\Title{Nuclear matter EOS with light clusters within the mean-field
	 approximation}

\bigskip\bigskip


\begin{raggedright}

{{\it M\'arcio Ferreira} and {\it Constan\c{c}a Provid\^encia}\\
 Centro de F\'isica Computacional\\ 
 Department of Physics, University of Coimbra\\ 
 P-3004-516 Coimbra\\ 
 Portugal\\ 
{\tt Email: mferreira@teor.fis.uc.pt, cp@teor.fis.uc.pt}}
\bigskip\bigskip
\end{raggedright}

\section{Introduction}

The crust of a neutron star is essentially determined by the low-density region 
($\rho<\rho_0\approx0.15-0.16\unit{fm}^{-3}$) of the equation of state.
At the bottom of the inner crust, where the density is $\rho\lesssim0.1\rho_0$, the formation of light clusters in 
nuclear matter will be energetically favorable at finite temperature. 
At very low densities
and moderate temperatures, the few body correlations are expected to become important and 
light nuclei like
deuterons  ($d\equiv\, ^{2}\text{H}$), tritons ($t\equiv\, ^3\text{H}$), helions
($h\equiv\, ^3\text{He}$) and $\alpha$-particles ($^4\text{He}$) will form. Due to Pauli blocking,
these clusters will dissolve at higher densities $\rho\gtrsim 0.1\rho_0$.
 The presence of these clusters influences the cooling process and
 quantities, such as the neutrino emissivity
and gravitational waves emission. The dissolution
density of these light clusters, treated as point-like particles, will be studied within the Relativistic Mean Field  
approximation. In particular, the dependence of the dissolution density on the clusters-meson couplings is studied \cite{Ferreira:2012ha}.
\section{The model}
The Lagrangian density of a system of nucleons and light clusters  ($d$, $h$, $t$, and $\alpha$ particles) is given by
\begin{equation*}
\mathcal{L}=\sum_{j=p,n,t,h}\mathcal{L}_{j}+\mathcal{L}_{{\alpha }}+
\mathcal{L}_d+ \mathcal{\,L}_{{\sigma }}+ \mathcal{L}_{{\omega }} + 
\mathcal{L}_{{\rho }} + \mathcal{L}_{\omega \rho},
\label{lag}
\end{equation*}
where the Lagrangian density $\mathcal{L}_{j}$ is
\begin{equation}
\mathcal{L}_{j}=\bar{\psi}_{j}\left[ \gamma _{\mu }iD^{\mu }_j-M^{*}_j\right]
\psi _{j}  \label{lagnucl},
\end{equation}
with 
\begin{eqnarray}
iD^{\mu }_j &=&i\partial ^{\mu }-g_v^j\omega^{\mu }-\frac{g_{\rho }^j}{2}{\boldsymbol{\tau}}%
\cdot \mathbf{b}^{\mu } 
, \label{Dmu} \\
M^{*}_j &=&M_j-g_{s}^j\sigma, \quad j=p,n,t,h
\label{Mstar}
\end{eqnarray}
where $M_{p,n}=M=938.918695\unit{MeV}$ is the nucleon mass and $M_{t,h}=A_{t,h}M - B_{t,h}$ are
the triton and helion masses. The binding 
energies of the clusters $B_{i}$ are in \cite{Ferreira:2012ha}. The couplings between the cluster $i$ and
the meson fields $\omega$, $\sigma$ and $\rho$ are given by $g_v^j$, $g_s^j$ and $g_\rho^j$, respectively.
The $\alpha$ particles and the deuterons are described as in \cite{typel2010}
\begin{eqnarray}
\mathcal{L}_{\alpha }&=&\frac{1}{2} (i D^{\mu}_{\alpha} \phi_{\alpha})^*
(i D_{\mu \alpha} \phi_{\alpha})-\frac{1}{2}\phi_{\alpha}^* \pc{M_{\alpha}^*}^2
\phi_{\alpha},\\
\mathcal{L}_{d}&=&\frac{1}{4} (i D^{\mu}_{d} \phi^{\nu}_{d}-
i D^{\nu}_{d} \phi^{\mu}_{d})^*
(i D_{d\mu} \phi_{d\nu}-i D_{d\nu} \phi_{d\mu})-\frac{1}{2}\phi^{\mu *}_{d} \pc{M_{d}^*}^2 \phi_{d\mu},
\end{eqnarray}
where $iD^{\mu }_j = i \partial ^{\mu }-g_v^j \omega^{\mu }$ and $M_{j}^*=M_{j}-g_s^j\sigma$ with $j=\alpha,d$.\\
The meson Lagrangian densities are 
\begin{eqnarray*}
\mathcal{L}_{{\sigma }} &=&\frac{1}{2}\left( \partial _{\mu }\sigma \partial %
^{\mu }\sigma -m_{s}^{2}\sigma ^{2}-\frac{1}{3}\kappa g_s^3 \sigma ^{3}-\frac{1}{12}%
\lambda g_s^4\sigma ^{4}\right)  \\
\mathcal{L}_{{\omega }} &=&\frac{1}{2} \left(-\frac{1}{2} \Omega _{\mu \nu }
\Omega ^{\mu \nu }+ m_{v}^{2}\omega_{\mu }\omega^{\mu }
+\frac{1}{12}\xi g_{v}^{4}(\omega_{\mu}\omega^{\mu })^{2} 
\right) \\
\mathcal{L}_{{\rho }} &=&\frac{1}{2} \left(-\frac{1}{2}
\mathbf{B}_{\mu \nu }\cdot \mathbf{B}^{\mu
\nu }+ m_{\rho }^{2}\mathbf{b}_{\mu }\cdot \mathbf{b}^{\mu } \right)\\
\mathcal{L}_{\omega \rho } &=& \Lambda_v g_v^2 g_\rho^2 \omega_{\mu }\omega^{\mu }
\mathbf{b}_{\mu }\cdot \mathbf{b}^{\mu }
\end{eqnarray*}
where $\Omega _{\mu \nu }=\partial _{\mu }\omega_{\nu }-\partial _{\nu }\omega_{\mu }$, 
$\mathbf{B}_{\mu \nu }=\partial _{\mu }\mathbf{b}_{\nu }-\partial _{\nu }
\mathbf{b}
_{\mu }-g_{\rho }(\mathbf{b}_{\mu }\times \mathbf{b}_{\nu })$. 

The equations of motion in the relativistic mean field approximation
 for the meson fields are given by 
\begin{align*}
 &m_s^2\sigma +\frac{\kappa}{2}g_s^3\sigma^2+\frac{\lambda}{6}g_s^4\sigma^3=\sum_{i=p,n,t,h}g_s^i\rho_s^i+\sum_{i=d,\alpha}g_s^i\rho_i\\
&m_v^2\omega^0+\frac{1}{6}\xi g_v^4\pc{\omega^0}^3+2\Lambda_v g_v^2g_\rho^2\omega^0\pc{b_3^{0}}^2=\sum_{i=p,n,t,h,d,\alpha}g_v^i\rho_i\\
&m_\rho^2b_3^{0}+2\Lambda_v g_v^2g_\rho^2\pc{\omega^0}^2b_3^{0}=\sum_{i=p,n,t,h}g_\rho^i I_3^i\rho_i 
\end{align*}
where $\rho_s^i=\frac{2S^i+1}{2\pi^2}\int_0^{k_f^i}k^2dk\frac{M_i-g_s^i\sigma}{\sqrt{k^2+\pc{M_i-g_s^i\sigma}^2}}$
is the scalar density, 
$\rho_i$ is the vector density, and $I_3^i$ ($S^i$) is the isospin (spin) of the specie $i$ for $i=n,p,t,h$. 
At zero temperature all the $\alpha$ and deuteron 
populations will condense in a state of zero momentum, therefore $\rho_\alpha$ and $\rho_d$ correspond to
the condensate density of each specie.
\section{Results}
\begin{figure*}
\hspace{-1.2cm}
\includegraphics[width=0.27\linewidth,angle=-90]{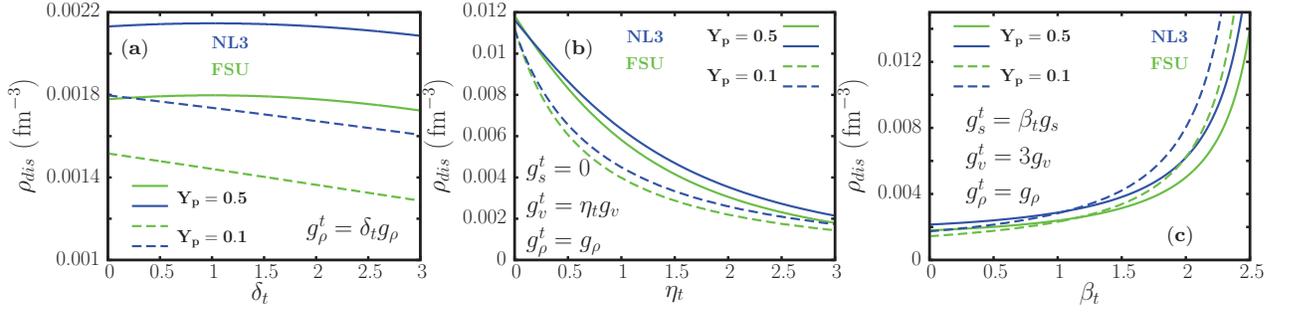}
\caption{The dissolution density at $T=0$ for
 tritons with $Y_p=0.5$ (solid lines) and $Y_p=0.1$ (dashed lines) as a function of: a) $\delta^t$ ($g_\rho^t=\delta_t g_\rho$), 
b) $\eta_t$ ($g_v^t=\eta_t g_v$) and c)$\beta_t$ ($g_s^t=\beta_t g_s$)   }
\label{fig:1}
\end{figure*}

%
The dissolution density $\rho_d$ of each light cluster is studied separately. We assume the coupling constants as $g_{v}^j=\eta_j g_v$,
$\,g_{s}^j=\beta_j g_s$, $\,g_{\rho}^i=\delta_i g_\rho$ and $g_{\rho}^l=0$ with $j=t,h,d,\alpha$, 
$i=t,\,h$ and $l=d,\,\alpha$ respectively. Then, we determine the dissolution density $\rho_d$ dependence
on the parameters $\eta_j$, $\beta_i$ and $\delta_i$ for different global proton fractions $Y_P$. The results for triton
at $T=0$ are shown in Fig. \ref{fig:1} (the results for all clusters can be seen in \cite{Ferreira:2012ha}).

The cluster formation and dissolution are insensitive to the $\rho$-meson-cluster
coupling constants when compared with $g_s^i$ and $g_v^i$ \cite{Ferreira:2012ha}. Therefore,
we fix the $g_\rho^i$ for triton and helion as $g_\rho$.  The $\rho_d$
 decreases with an increase of the $\omega$-meson-cluster coupling constant and increases
 with an increase of the $\sigma$-meson-cluster coupling constant. For some value of $\beta_i$ below the mass number
$A_i$, $\rho_d$ increases abruptly and the clusters will not dissolve.

To fix $g_s^i$ and $g_v^i$ for each cluster two constraints are required.
We  consider  the 
$\rho_d$ for symmetric nuclear matter ($Y_p=0.5$) {at $T=0$ MeV} \cite{ropke2009}, the in-medium 
binding energies $B_i=A_i M^*-M^*_i$ \cite{typel2010}, and recent experimentally Mott points at $T\approx5$ MeV \cite{hagel2012}.
\begin{figure*}[h]
\includegraphics[width=0.33\linewidth,angle=-90]{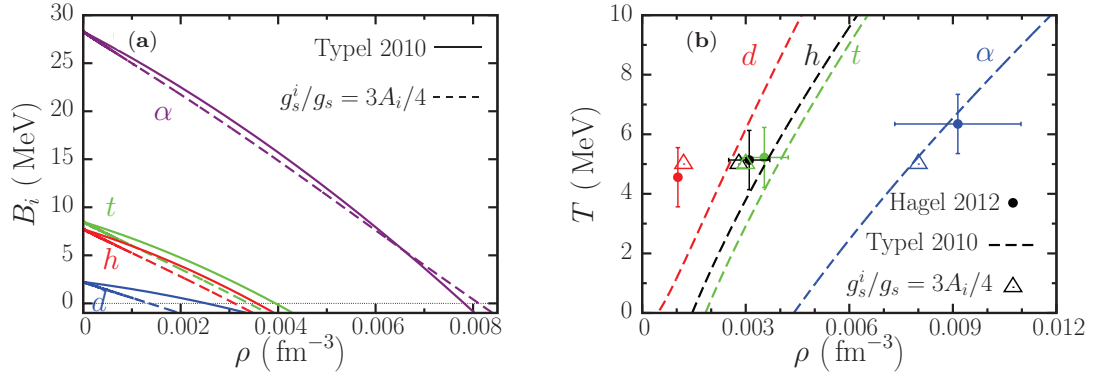}
\caption{The in-medium binding energy in our model is compared in (a), for all the clusters (dashed lines), with the corresponding
results of Typel 2010 \cite{typel2010} at $T=5$ MeV (full lines). In (b)  we compare our Mott points with the ones from Typel 2010 \cite{typel2010} 
(dashed lines) and the experimental prediction by Hagel 2012 \cite{hagel2012} (full dots with
errorbars). The FSU parametrization was considered.}
\label{fig:2}
\end{figure*}
We contruct several sets with $g_s^i/g_s=0,\,A_i/2,\,3A_i/4$ and $\,4A_i/5$, and calculate the $g_v^i/g_v$ value 
of each set that reproduces $\rho_d$ for $Y_p=0.5$ { at $T=0$ MeV} \cite{ropke2009}. After, we compare
the behavior of the different sets at $T=5\,\mathrm{MeV}$ with the in-medium 
binding energies \cite{typel2010} and the Mott points \cite{hagel2012}. The results for the best parameter 
set \cite{Ferreira:2012ha} are shown in Figure \ref{fig:2}. To test the parametrization chosen, we calculate the particle 
fraction at finite temperature and in chemical
equilibrium, $\mu_i=Z_i \mu_p+(A_i-Z_i) \mu_n$ with $i=\alpha, \, h,\,t, \, d$ \cite{alphas,clusters}.
\begin{figure*}[h]
\hspace{-0.8cm}
\includegraphics[width=0.25\linewidth,angle=-90]{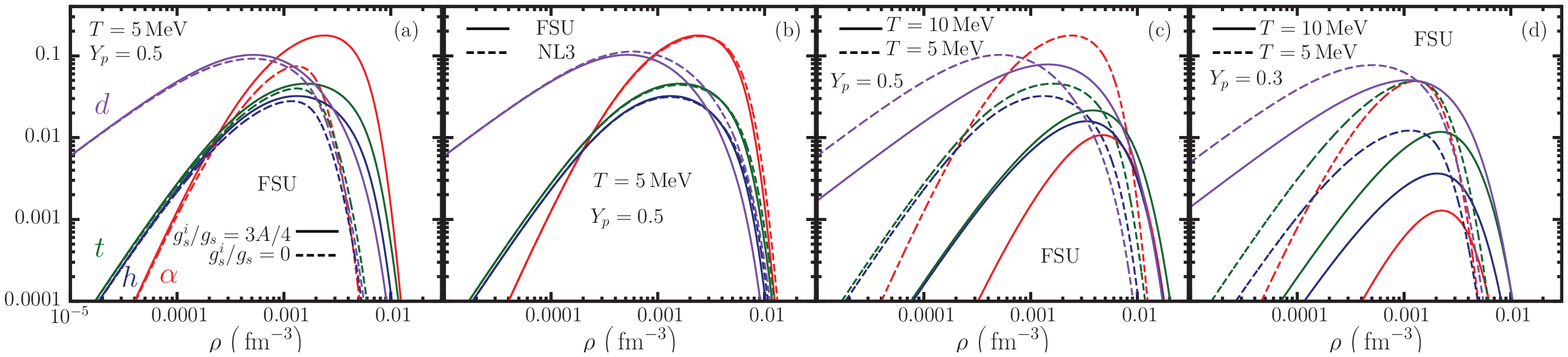}
\caption{Fraction of light clusters in equilibrium with nuclear matter: (a) 
 for FSU with  $Y_p=0.5$ and $T=5$ MeV, and $g_{s}^i=0$ (dashed lines) and
 $g_{s}^i=3\, A_i/4\, g_s$ (full lines); (b)  for $g_{s}^i=3\, A_i/4\, g_s$ with
 $Y_p=0.5$ and $T=5$ MeV  and FSU (full lines) and NL3 (dashed  lines); (c)
 for FSU with   $g_{s}^i=3\, A_i/4\, g_s$ and $Y_p=0.5$, and $T=10$ MeV (full
 lines) and $T=5$ MeV  (dashed  lines);  (d)
 for FSU with   $g_{s}^i=3\, A_i/4\, g_s$ and $Y_p=0.3$, and $T=10$ MeV (full
 lines) and $T=5$ MeV  (dashed  lines).}
\label{fig:3}
\end{figure*}
The effect of $g_s^i$ at $T=5$ MeV is shown in Fig. \ref{fig:3} (a). In Fig. \ref{fig:3} (b) -- (d), we always set $g_s^i= 3A_i/4\, g_s$.
In Fig.\ref{fig:3} (b) we compare two models, NL3 (dashed lines) and FSU
(full lines).
 The effect of temperature and isopsin asymmetry is
shown in Fig.\ref{fig:3} (c) and (d). In
Fig.\ref{fig:3} (c) particle fractions in symmetric matter are compared for $T=10$ MeV (full lines) and  $T=5$ MeV (dashed lines).\\
For a larger coupling $g_s^i/g_s$, both the fraction of particles and the dissolution density are larger.
For symmetric nuclear matter at $T=10$ MeV, the deuteron
fraction is already the largest fraction and the $\alpha$ fraction the
smallest. In neutron rich matter, the deuteron fraction is the largest and the tritons come
in second both at $T=5$ and $10$ MeV. The results do not depend much on the nuclear parametrization chosen.
\section{Conclusion}
Light clusters have been included in the EOS of nuclear matter as point-like
particles with constant coupling constants within relativistic mean field approach.
Results demonstrate that the dissolution of clusters is mainly determined by the
isoscalar part of the EOS. Recent experimental results for the in-medium binding energy
of light clusters \cite{hagel2012} at $T\sim 5$ MeV and results from a
quantum statistical approach \cite{typel2010} were used to constraint the clusters
couplings constants.
It was shown that a
larger $\sigma$-cluster coupling gives rise to larger dissolution densities
and larger particle fractions in chemical
equilibrium at $T=5\unit{MeV}$ and $T=10\unit{MeV}$ for
symmetric and asymmetric nuclear matter with light clusters. 

The in-medium binding energies proposed in \cite{typel2010} may
be reproduced  within our model with a temperature dependent meson-cluster coupling constants.
More experimental Mott points at more temperatures would allow the determination of the
temperature dependence. 
\section{Acknowledgments}
This work was partially supported by QREN/FEDER, through the Programa Operacional Factores de Competitividade - COMPETE
and by National funds through FCT - Funda\c{c}\~{a}o para a Ci\^{e}ncia e Tecnologia under Project No. PTDC/FIS/
113292/2009 and Grant No. SFRH/BD/51717/2011.

\end{document}